\documentclass[prl,twocolumn,notitlepage,amsfonts,amssymb,amsmath,floatfix,tightenlines,superscriptaddress]{revtex4-2}

\usepackage{amsfonts,amssymb}
\usepackage{amsmath,times}
\usepackage{graphicx}
\usepackage[colorlinks=true,linkcolor=blue,anchorcolor=red,citecolor=blue,urlcolor=blue]{hyperref}
\usepackage{color}
\usepackage{cleveref}
\usepackage[lofdepth,lotdepth,caption=false]{subfig}

          \usepackage[normalem]{ulem}                      

\usepackage[colorlinks=true,linkcolor=blue,anchorcolor=red,citecolor=blue,urlcolor=blue]{hyperref}

\begin{document}

\title{Kekul\'e Superconductivity in Twisted Magic Angle Bilayer Graphene}

\author{Ke Wang}
\affiliation{Department of Physics and James Franck Institute, University of Chicago, Chicago, Illinois 60637, USA}
\affiliation{Kadanoff Center for Theoretical Physics, University of Chicago, Chicago, Illinois 60637, USA}
\author{K. Levin}
\affiliation{Department of Physics and James Franck Institute, University of Chicago, Chicago, Illinois 60637, USA}

\date{\today}


\begin{abstract}
While it has been one of the most important new physics
discoveries in the last decade, the nature of
superconductivity in the twisted graphene
family remains an unsolved problem.
Motivated by recent scanning tunneling experiments that report Kekul\'e
ordering in moir\'e graphene superconductors, we develop a microscopic
theory of this superconductivity for the twisted bilayer system. The
pairing we find is an intra-valley, finite-momentum pair-density wave
(PDW) that intrinsically carries a Kekul\'e modulation. This state
exhibits four salient features: (i) spontaneous breaking of $C_3$
rotation symmetry, producing nematic order (ii)with triplet pairing;
and (iii) a
quasiparticle density of states that evolves from a V-shaped profile to
a fully gapped, U-shaped spectrum as the attraction increases which is
accompanied by (iv) systematic behavior of the temperature dependent zero
bias conductance. These features align with key experimental signatures.
We find, as well, that with only modest interaction strengths, the
state is near to a BEC-like phase, consistent with the observed
extremely short coherence lengths. Taken together, these results
identify a microscopic intra-valley Kekul\'e PDW as a compelling
candidate for unconventional superconductivity in the twisted graphene
family.
\end{abstract}

\maketitle

\section{Introduction} 

It is hard to overestimate the excitement which has been generated by the
discovery~\cite{Cao2018a} of ``high" temperature superconductivity associated with
the flat band regime~\cite{Bistritzer2011} in twisted bi- and tri-layer graphene~\cite{Nuckolls2024}.  Measured
so far are
transport properties \cite{Polshyn2019,Cao2020}, superfluid density\cite{Banerjee2025,Tanaka2025},
tunneling\cite{Jiang2019,Oh2021,Xie2019},  
upper critical fields~\cite{Cao2018a,Park2021}, thermodynamical~\cite{Zondiner2020}
as well as other superconducting characteristics\cite{Yankowitz2019,Sharpe2019,Stepanov2019,Saito2019}.

Among these, scanning tunneling spectroscopy and microscopy~\cite{Nuckolls2023,Kim2023}
have led to progress in understanding the nature of the correlated
insulators.
What is most remarkable is that evidence for an inter-valley coherent (IVC) ordering,
called Kekul\'e was anticipated theoretically~\cite{Wagner2022,Kwan2021}. This appears
at $ \nu = \pm 2$ and
$\pm 3$ 
which
coincides with a $\sqrt{3} \times \sqrt{3}$ reconstruction of the atomic scale unit cell.
While there has been much theoretical attention~\cite{Wagner2022,Kwan2021} paid to the implications for
the correlated insulator
regimes\cite{Nuckolls2023,Kim2023}, 
a form of Kekul\'e ordering has also been observed within the superconducting and pseudogap
phases in the same interval
($ \nu = \pm 2$ to counterpart
$\pm 3$.). 
This, we argue here, 
is essential to address in order to understand the superconductivity in these
twisted graphene, flatband systems.

What has been emphasized in the experimental literature is that 
(i) the superconductor arises out of a Kekul\'e order that is
qualitatively different from that of the neighboring insulator\cite{Nuckolls2023},
(ii) that the superconducting and pseudogap phases look similar
to one another with, thus far, no readily distinguishable features in imaging 
\cite{Nuckolls2023}. (iii) Importantly,
the strength of the IVC order, quantified by the normalized Kekul\'e peak
intensity is maximum at filling factors corresponding to those with
very strong pairing, as evidenced by the short coherence lengths
and the presence of a pseudogap phase\cite{Kim2023}.

\begin{figure}[h]
    \centering
    \includegraphics[width=3.6in]{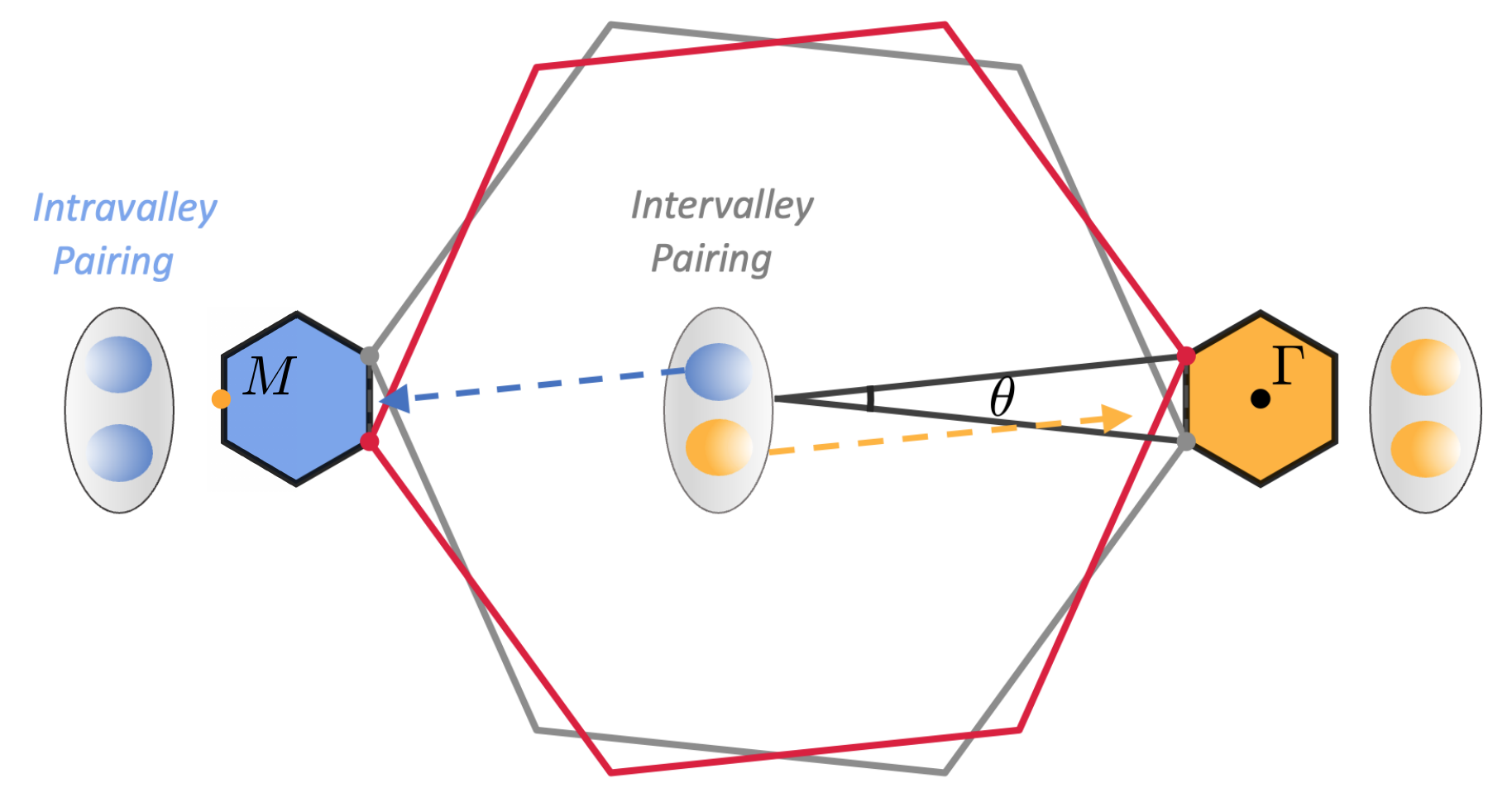}
    \caption{Cartoon of the twisted bilayer graphene geometry and Cooper Pairs. This serves to contrast intravalley
and intervalley pairing. Two graphene Brillouin zones (BZs) are rotated by $\pm\theta$ (gray/red). Two mini-moiré BZs
associated with
$K/K'$ valleys are represented by orange and blue;  the $\Gamma/M$-point is shown in the right/left mini-BZ.
Intravalley pairing involves pairs {\it within} each valley and associated mini-BZ, while intervalley pairing
involves two different mini-BZs. }
\label{fig1}
\end{figure}

Motivated by this STM data, we now consider intra-valley pairing superconductivity in twisted bilayer graphene (TBG), which is a methodology for implementing superconducting Kekul\'e order~\cite{Roy2010,Tsuchiya2016,yafis2025}. 
We will see that one can alternatively view this intravalley pairing as a
pair density wave (PDW) in which the pairs have non-zero net momentum.
In this way, it should be emphasized that ``Kekul\'e superconductivity" as discussed
here refers not
to a Kekul\'e pairing mechanism or ``glue" but to a Kekul\'e pairing machinery.

In a qualitative way we can now understand the STM experiments. From (i) it would then appear that one is
seeing two distinct Kekul\'e
orders that live in different sectors:
in the insulator, a particle-hole Kekul\'e (bond/charge/valley-coherent) order, and
in the superconductor, a particle-particle Kekul\'e component.
From (ii) the
observation that pseudogap and superconducting phases
show (essentially) the same Kekul\'e pattern fits naturally with a Kekul\'e
driven pseudogap containing the same preformed pairs as those
that become condensed within the superconducting phase.
From (iii) that the highest Kekul\'e peaks are associated with the
strongest pairing
seems to strongly support the idea that Kekul\'e ordering and the
superconducting pairing machineries are
correlated \footnote{To be precise, this correlation should not be viewed as a
``Kekul\'e-driven pairing mechanism''. As shown in arXiv:2506.18996,  
quite generally, the strongest pairing tends to occur where the non-superconducting
ordering (in this case, Kekul\'e) is weakest, as in a quantum-critical-point scenario
and also more generally.}.

This analysis underlines the importance of systematically
addressing, as we do here, the intra-valley
pairing case which leads to Kekule superconductivity.
This proposal stands in contrast to the majority of current theories, which focus on inter-valley pairing~\cite{You2019,Wu2018,Lian2019,Cea2021,Hu2019,Julku2020,Christos2023} (and its variants~\cite{Khalaf2021,Gonzalez2019,Po2018}).
The simple representation in Figure~\ref{fig1} emphasizes the differences in these two scenarios.
One cannot, however, rule out that inter-valley and intra-valley pairing may coexist; 
they originate from distinct, orthogonal symmetry channels of the underlying microscopic 
attraction. 

We emphasize throughout that this
intra-valley pairing scenario leads to important and distinct theoretical consequences from the
inter-valley case, some of
which are rooted in the structure of the flat-band Bloch wavefunctions. The presence of 
an effect
which we term the ``quantum textures" 
emphasizes a notable contrast. Unlike conventional inter-valley pairing where details
about the wavefunction are generally absent from the gap equation, this quantum texture,
containing wavefunction information and multi-band effects,
enters directly into the intra-valley gap equation. It strongly modulates the behavior of 
the order parameter. We will see the interplay between this texture and the fundamental symmetries of a single moir\'e valley dictates the nature of the resulting superconducting order.

\section{PDW Order and Microscopic Model}
A pair density wave superconductor~\cite{Agterberg2020}
involves pairs with finite center-of-mass momentum, $2\mathbf{Q}$. The order is bond-centered, forming on the AB/BA sublattice bonds, and in the low-filling regime, \textit{the pairing momentum $\mathbf{Q}$ is naturally expected to lie near the Dirac point ${\bf K}$.} The full order parameter is a time-reversal-symmetric combination of condensates from the two valleys (e.g., $[\Delta_{2\mathbf{Q}},\Delta_{-2\mathbf{Q}}]$).

\begin{figure}
    \centering
    \includegraphics[width=\linewidth]{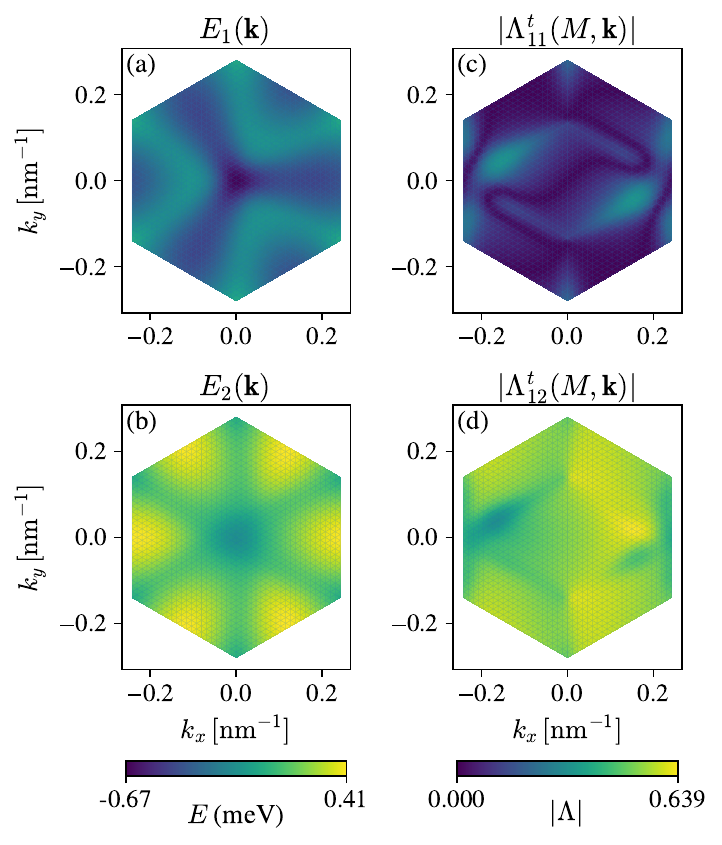}
   \caption{Flat-band dispersion and form factors from the Bistritzer--MacDonald (BM) model in the mini-Brillouin zone. Parameters: $w_{AA}=0.08$ eV, $w_{AB}=0.11$ eV, twist angle $\theta\simeq 0.94$, and $\hbar v_F=0.684$ eV$\cdot$nm. (a,b) Band energies $E_1(\mathbf{k})$ and $E_2(\mathbf{k})$ measured from half-filling; each flat band has bandwidth 
$W \approx 0.5$ meV and respects three-fold rotation $C_3$. (c,d) Magnitudes of the intra-band and inter-band form factors (taken from layer $L=1$) with pairing momentum $\mathbf{Q}=M$, which enter into the gap equation and are central to superconductivity. Note that in the triplet channel, $C_2\mathcal{T}$ symmetry strongly suppresses intraband (diagonal) pairing, whereas interband (off-diagonal) pairing remains strong; consequently, the pairing is dominated by interband form factors.
} 

    \label{fig2}
\end{figure}
Another crucial constraint on the nature of this PDW superconductivity
comes from the fact that 
STM experiments cannot probe a PDW directly but rather are sensitive to its
consequences through the charge sector which must reveal the specific
form of Kekul\'e order.
One important consequence of these two-valley PDW states is that they can induce secondary
order, involving charge-density waves (CDWs). A CDW arises from the bilinear coupling of the
two valley condensates,
\begin{equation}
  \rho(\mathbf r)\;\propto\;
  \Delta_{2\mathbf Q}(\mathbf r)\,\Delta^{*}_{-2\mathbf Q}(\mathbf r)
  + \text{h.c.},
\end{equation}
which carries total momentum $4\mathbf Q$. On the atomic lattice this reduces to a
modulation $\propto \cos(2\mathbf K\!\cdot\!\mathbf r)$ (since $3\mathbf K$ is a
reciprocal vector), yielding a Kekul\'e $\sqrt{3}\!\times\!\sqrt{3}$ superlattice,
i.e., a tripled unit cell. The slowly varying moir\'e-scale envelope is set by
$
\boldsymbol{\delta q}=2\mathbf Q-2\mathbf K
$
and thus varies on the length scale $|\boldsymbol{\delta q}|^{-1}$. Although a full
analysis of this secondary order is beyond the scope of this work, this mechanism
provides a connection between the proposed PDW and the STM experiments~\footnote{
A coexisting intervalley condensate with zero center-of-mass momentum, $\Delta_0$, can 
also couple to the PDW order to induce a secondary CDW:
The charge modulation from this second mechanism has a momentum of $2\mathbf{Q}$ and, like the first, behaves as $\cos(2\mathbf{K}\cdot\mathbf{r})$ on the atomic scale.
This will then lead to an observed Kekule pattern in the charge channel which can
be observed in STM.
}.

Although intra-valley pairing takes place 
in two different valleys, it is sufficient for our purposes to analyze
a single valley separately. That the two valleys contribute independently is
well-justified for small twist angles. 
%
Our microscopic theory is based on the Bistritzer-MacDonald (BM) continuum model for TBG\cite{Bistritzer2011,PhysRevB.100.035448,PhysRevB.99.035111}, supplemented with a finite, but short range attractive interaction.

The BM continuum model captures the essential moir\'e physics through two interlayer tunneling parameters, $w_{AA}$ and $w_{AB}$, together with the twist angle $\theta$. Since the flat bands are not rigid and their dispersion can vary with BM parameters and interaction effects, we test the robustness of our superconducting solution across a range of flat-band bandwidths. In practice, we fix the relaxation-renormalized tunneling parameters to $w_{AA}=80~\mathrm{meV}$ and $w_{AB}=110~\mathrm{meV}$, and tune the twist angle in the vicinity of the magic angle so that the bandwidth varies from $\sim 1~\mathrm{meV}$ to $\sim 10~\mathrm{meV}$~\footnote{In the metallic (fractional-filling) regime relevant to superconductivity, the long-range tail of the Coulomb repulsion is expected to be strongly screened. In this situation the residual screened interaction is relatively weakly momentum dependent at small $\mathbf{q}$ and primarily renormalizes the low-energy dispersion. Varying the bandwidth therefore provides a practical way to test the robustness of our conclusions against generic modifications of the bandstructure.}.

A generic translationally invariant pairing interaction can be written as
\begin{equation} \begin{split} \hat{V} = \sum V_{aa'}(\mathbf{q}-\mathbf{q}')\, &\psi^\dagger_{a,\sigma}(\mathbf{k}_+)\psi^\dagger_{a',\sigma'}(\mathbf{k}_-)\, \\ &\times \psi_{a',\sigma'}(\mathbf{k}'_-)\psi_{a,\sigma}(\mathbf{k}'_+), \end{split} \end{equation}
where $\mathbf{k}_\pm=\mathbf{k}\pm\mathbf{q}$ and $\mathbf{k}'_\pm=\mathbf{k}'\pm\mathbf{q}'$. The indices $\sigma,\sigma'$ label spin, and $a,a'$ correspond to the layer/sublattice. We require the interaction to respect the symmetries of the single-valley BM model and, consistent with our focus on AB-bonding order, to couple different sublattices.

To keep the analysis sufficiently generic, we consider a class of short-range attractive interactions satisfying these criteria. Specifically, within each layer we take an attraction between sites at $\mathbf{r}$ and $\mathbf{r}+\boldsymbol{\delta}_i+\mathbf{d}$, where $\boldsymbol{\delta}_i$ are nearest-neighbor vectors of the honeycomb lattice and $\mathbf{d}$ is a graphene lattice translation vector with $|\mathbf{d}|\ll a_M$, with $a_M$ the moir\'e lattice constant. In practice, we vary $|\mathbf{d}|$ up to $\sim 10$ times the graphene lattice constant $a$. The commonly used nearest-neighbor (NN) attraction corresponds to $\mathbf{d}=0$ (see, e.g., Refs.~\cite{Roy2010,Tsuchiya2016}). Screening effects~\cite{Goodwin2019,Pizarro2019} are important at fractional filling in twisted graphene\cite{Stepanov2020,Saito2020}, and the repulsive core of the screened Coulomb interaction is estimated to extend over only a few graphene lattice constants (e.g., $\sim 5a$). This motivates our focus on non-local short-range attractions with 
$|\mathbf{d}| \ll a_m$
for which the dominant on-site/ultrashort-range repulsion is effectively avoided.

In this work, we do not aim to determine the microscopic origin of the pairing glue; instead, we focus on the nature of the superconducting order in the presence of such a generic short-range, non-local attractive interaction. Microscopically, the attraction could arise from electron--electron interactions mediated by the exchange of bosonic fluctuations (e.g., spin or valley fluctuations~\cite{You2019}). See Sec.~{\it ``Overview of Pairing Theories ''} for additional discussion.

We have numerically checked that the essential properties of the resulting superconducting state remain robust upon varying the bandwidth and the interaction range within the range of physical constraints. We thus view our main conclusions as generic and not tied to the microscopic details of the assumed short-range attractive interaction.
For definiteness, the numerical results and figures shown in this work are obtained using the simple nearest
neighbor (NN) electron-electron interaction model.

%

\section{Order Parameter and Quantum Textures}

We begin by utilizing the spatial $C_3$ rotational symmetry of the lattice to classify the possible superconducting pairing channels. The $C_3$ point group has three one-dimensional irreducible representations, which we label as $A$, $E_1$, and $E_2$. An eigenstate belonging to one of these irreducible representations
acquires a phase of $\zeta^l,\, l=0,1,2$  under a $C_3$ rotation, where $\zeta = \exp(i2\pi/3)$. The interaction potential can be decomposed into these eigenstates. For a generic short-range attraction between sites at $\mathbf{r}$ and $\mathbf{r} + \boldsymbol{\delta}_i + \mathbf{d}$,, we have the decomposition for each layer:
\begin{equation} V(\mathbf{q}-\mathbf{q}')= \sum_{l=0}^2 g_l f_l(\mathbf{q}) f_l^*(\mathbf{q}'). \label{2} \end{equation} 
Here $f_l(\mathbf{q})=3^{-1}\sum_{j=0}^2 \zeta^{jl}\, \exp {-i\mathbf{q}\cdot({\bf d}+\boldsymbol{\delta}_j}) $.
As shown in Figure 1, in the case of intravalley pairing, Cooper pairs consist of electrons originating from the same moiré valley. Given that the mini-Brillouin zone is small and $\mathbf{G}$-components of flat-band wavefunctions are limited to the first few Umklapp vectors, the relevant momenta $|\mathbf{q} + \mathbf{G}|$ remain significantly smaller than any atomic-scale momentum scale. In this limit, the form factor for the $A$-channel ($l=0$) approaches $f_0(\mathbf{q}+ \mathbf{G}) \to 1$, whereas the chiral channels ($l=1, 2$) vanish as $f_{1,2}(\mathbf{q}+ \mathbf{G}) \to 0$. Thus, the $A$ channel emerges as the dominant superconducting symmetry. This analysis remains general for $|\mathbf{d}| \ll a_m$, providing a fundamental explanation for the robustness of superconductivity across a broad class of short-range models.
 
\begin{figure}
    \centering
    \includegraphics[width=\linewidth]{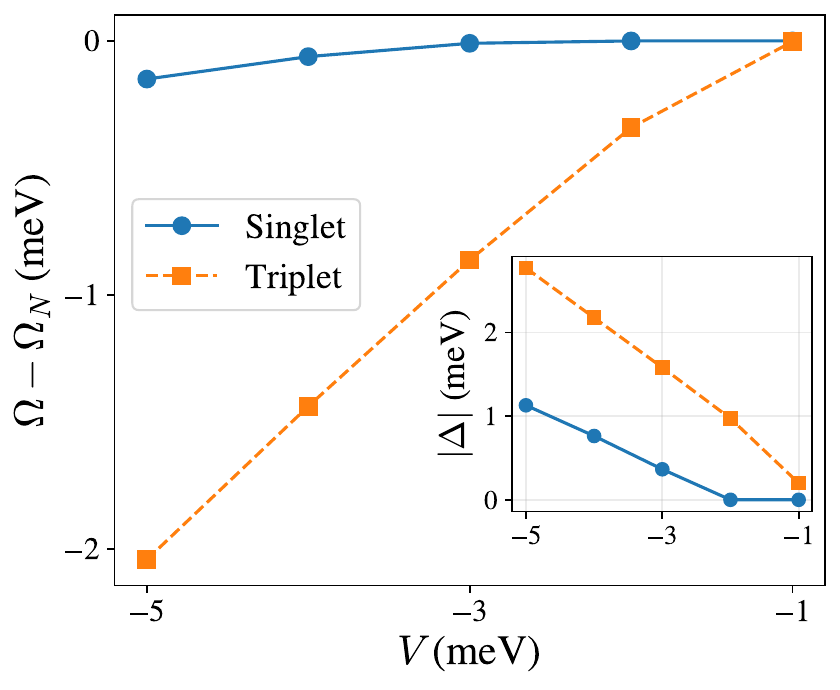}
\caption{
The figure shows the grand-canonical thermodynamic potential $\Omega$ of the singlet and triplet pairing states as a function of interaction strength $V \equiv V_0(q=0)$. Results are computed at near-zero temperature $T=0.001~\mathrm{meV}$ and chemical potential $\mu=-0.59~\mathrm{meV}$. The plotted quantity is $\Omega-\Omega_N$, measured relative to the normal-state value $\Omega_N$. The inset shows the corresponding superconducting order-parameter magnitude $|\Delta|$ for the two states. For the interaction strengths shown, the triplet state has a lower $\Omega$ than the singlet state, indicating that triplet pairing is thermodynamically favored.
}
\label{fig3}
\end{figure}

For this case which we consider throughout the paper, the pairing gap parameter satisfies:
\begin{equation}
    {\Delta}_{ aa',\sigma\sigma'}(\mathbf{Q}) = - \sum_{\mathbf{q}} V_{0}(\mathbf{q})\, \langle \hat{\psi}_{a',\sigma'}( \mathbf{Q}_-) \hat{\psi}_{a,\sigma}(\mathbf{Q}_+) \rangle. \label{3}
\end{equation}
Here $V_{0}(\mathbf{q})= g_0 f^*_{0}(\mathbf{q})$
and $\mathbf{Q}_\pm = \mathbf{Q}\pm \mathbf{q}$.
In the NN model, we have the layer/sublattice indices $a=LA$ and $a'=LB$. This is consistent with
the physics of a condensate which forms on the AB sublattice bonds within each layer. Furthermore, the condensates in the two layers are related by mirror symmetry such that they are equal in magnitude.

To analyze the pairing further, we rewrite the order parameter in the spectral (or band) basis. The transformation from the plane-wave basis (operators $\hat{\psi}$) to the spectral basis (operators $c$) requires the introduction of the Bloch wavefunctions, $u_{\mathbf{G},a;n}(\mathbf{k})$:
\begin{equation}
    \hat{\psi}_{a,\sigma}(\mathbf{k}+\mathbf{G}) = \sum_n u_{\mathbf{G},a;n}(\mathbf{k})\, \hat c_{n,\sigma}(\mathbf{k})
\end{equation}
This leads to the important definition of a (particle-particle) form factor, $\Lambda_{mn,aa'}(\mathbf{k},\mathbf{q})$, which projects the interaction onto the particle-particle channel in this bandstructure basis:
\begin{equation}
    \Lambda_{mn,aa'} = \sum_{\mathbf{G}} f^*_{0}(\mathbf{q}+\mathbf{G})\, u_{\mathbf{G},a;m}(\mathbf{Q}_+)\, u_{-\mathbf{G},a';n}(\mathbf{Q}_-). \label{5}
\end{equation}
Here the $A$-channel $f_0$ factor depends on the form of the interaction. As analyzed below Eq.~\ref{2}, for generic short-range interactions $f_0$ is close to a constant. This indicates that the form factor is primarily controlled by the BM-model wavefunctions participating in the 
pairing. It is largely insensitive to microscopic details of the attraction.

  Using this form factor, Eq.~\ref{3} can be rewritten in terms of:
\begin{equation} 
{\Delta}_{ aa',\sigma\sigma'} =g_0 \sum  \Lambda_{mn}(\mathbf{Q},\mathbf{q})\, \langle  \, \hat c_{m,\sigma}(\mathbf{Q}_+)  \hat c_{n,\sigma'}( \mathbf{Q}_-) \rangle. \label{6}
\end{equation}
Here the summation is over the band index $m/n$ and momentum ${\bf q}$ in the mini-BZ. 
We emphasize that in the form factor of Eq.~\ref{5}, only direct \emph{products} of Bloch wavefunctions appear. This is to be contrasted with their \emph{inner products}
which would appear in the particle-hole sector. Time-reversal symmetry relates opposite valleys and is absent within a single moiré valley. Consequently, the 
diagonal form factor $\Lambda_{nn}(\mathbf{Q}, \mathbf{q}=0)$ generally 
deviates from unity, and the off-diagonal components $\Lambda_{mn}$ 
acquire sizeable magnitudes 
that are central to the pairing machinery. 
This structure, which reflects the unique quantum texture of the 
intra-valley sector, is addressed more explicitly in Fig.~\ref{fig2}. This is a key distinction between intra-valley pairing and BCS-like inter-valley pairing (or the correlated Kekulé insulator), where
the analogue diagonal form factors are close to unity.

In our numerical simulations, we truncate the remote bands to those nearest in energy to the flat bands. We find that the isolated-flat-band result ($N=2$) is already close to the multiband solution; for example, $\Delta(N=2)/\Delta(N=20)\approx 0.8$. We attribute this proximity
to the size of the energy gap separating the flat bands from higher-lying remote bands (see Methods, ``Flat-band approximation''). We have further checked $N=10,20,30$ and find that the qualitative superconducting properties remain robust. Remote bands play a more important role in the superfluid stiffness of Kekul\'e superconductivity \cite{Wang2026Superfluid}.

The pairing physics is governed by the momentum- and band-dependent form factor $\Lambda_{mn}(\mathbf{k})$ in the band basis. We refer to $\Lambda_{mn}(\mathbf{k})$ as a ``quantum texture,'' since it is determined by the moir\'e Bloch wavefunctions and their internal (Umklapp) structure, and it encodes how pairing weight is distributed in momentum and band space. This texture is a distinctive feature of moir\'e systems and plays a central role in our analysis: it controls the relative strength of competing pairing channels and underlies key qualitative properties of the condensate discussed in the next section. The same quantum texture also gives rise to a distinctive geometric electromagnetic response characteristic of Kekul\'e superconductivity \cite{Wang2026Superfluid}. 

Note that this texture does not arise in the same form in prior PDW-related theories for untwisted crystalline systems, such as the monolayer TMD setting of Ref.~\cite{aat4698} and the rhombohedral multilayer graphene setting of Ref.~\cite{PhysRevB.111.174523}, where the pairing interaction is typically represented by a simpler vertex rather than a moir\'e-wavefunction-derived texture.

\subsection{Establishing the Thermodynamically Stable Phase}

For a given pairing momentum ${\bf Q}$, one can solve the gap equation self-consistently to find possible solutions. Because not all solutions represent thermodynamically stable states it is necessary to investigate their stability by
establishing that they correspond to a global minimum of the grand-canonical potential $\Omega$ at a given chemical potential $\mu$. 

In the PDW state under consideration, the thermodynamic potential is a function $\Omega(\mu, {\bf Q}, \Delta_L, {\bf S})$, where we label the layer index $L=1,2$ and $\mathbf{S}$ represents the pair spin. There are two branches of solutions satisfying the gap equation: $\Delta_1=\Delta_2$ and $\Delta_1=-\Delta_2$. Both branches respect mirror symmetry. The first branch,
which is the physical solution, corresponds to a global minimum for a fixed ${\bf Q}$ and $\mu$, while the second branch represents a saddle point. Thus, throughout the paper we use $\Delta\equiv \Delta_1$ to denote the order-parameter magnitude.

In the stable ground state with $\Delta_1=\Delta_2$, the pairing momentum ${\bf Q}$ is treated as a \emph{variational} parameter constrained to lie within the mini-Brillouin zone (mBZ) \footnote{
The zero-current condition is automatically satisfied once the time-reversal partner in the opposite valley is included. This differs from the conventional Fulde--Ferrell (FF) state, where ${\bf Q}$ must be determined self-consistently to enforce zero current. Moreover, a conventional FF state is often associated with a saddle point of the thermodynamic potential (see, e.g., Wang \emph{et al.}, Phys.\ Rev.\ B \textbf{97}, 134513), whereas in multiband/multicomponent settings additional degrees of freedom can stabilize finite-${\bf Q}$ pairing.}. Note that the compactness of the mBZ guarantees that $\Omega({\bf Q})$ attains a minimum within the mBZ, in contrast to untwisted continuum systems where ${\bf Q}$ is not restricted to such a compact domain. 

For a given chemical potential $\mu$ and spin configuration (singlet or unitary triplet), we find that the thermodynamic potential $\Omega$ is minimized at ${\bf Q}=M$ within the mBZ. This provides a direct, experimentally testable prediction: the induced charge modulation contains a corresponding finite-$\mathbf{q}$ component, and in a layer-resolved description it appears as $4(M-K_L)$, where $L=0,1$ labels the top/bottom layer Dirac momentum. We emphasize that this corresponds to a moir\'e-scale modulation. In particular, the PDW mechanism yields a finite-$\mathbf{q}$ mini-BZ signature, in contrast to $K$-intervalley-coherent (KIVC) order, which produces a $\mathbf{q}=0$ mini-BZ signal in strain-free correlated insulators~\cite{PhysRevX.10.031034}.

We note that the relative thermodynamic potential $\Omega$ of the singlet and triplet solutions depends on the pairing momentum ${\bf Q}$. Importantly, when ${\bf Q}$ is at the $M$ point, the unitary triplet state has a lower $\Omega$ than the singlet. This is consistent with experimental indications of non-singlet pairing~\cite{Cao2021}. The singlet--triplet comparison is shown in Fig.~\ref{fig3}. In our case the triplet channel is {unitary/unpolarized} and is driven by strong interband pairing form factors.   This behavior contrasts with the \emph{spin-polarized} triplet state discussed for intravalley pairing in rhombohedral graphene~\cite{PhysRevB.111.174523} and inter-valley pairing considered in Ref.~\cite{Christos2023}, which is inherited from the spin-polarized normal state.

Since the normal-state Hamiltonian is spin-$SU(2)$ invariant (and we work at zero magnetic field), we may fix a spin quantization axis and represent any \emph{unitary} spin-triplet order parameter in the $S_z=0$ basis without loss of generality. With this convention, the singlet--triplet distinction appears through the spin matrix structure of the gap, while the remaining information relevant for our discussion is encoded in the form factors:
\begin{equation}
    \Lambda^{s/t}(\mathbf{Q}, \mathbf{q})
    =\frac{1}{2} \left[ 
     \Lambda(\mathbf{Q}, \mathbf{q}) \pm \Lambda^{\mathsf T}(\mathbf{Q}, -\mathbf{q}) \right] .
    \label{11}
\end{equation}
Here, $s$ and $t$ denote singlet and triplet, respectively, and the transpose acts on the band indices. Due to $C_2 \mathcal{T}$ symmetry, the diagonal component of $\Lambda^t$ (triplet channel) is strongly suppressed compared to the singlet channel\footnote{The $C_2 \mathcal{T}$ symmetry enforces the condition $\Lambda^{t}_{nn}(\mathbf{Q},\mathbf{q}=\mathbf{0})=0$.}. However, there are two active flat bands and a few remote bands near the chemical potential, and the off-diagonal pairing\cite{Christos2023} plays a crucial additional role here regarding the competition between singlet and triplet states which depend on ${\bf Q}$.

\begin{figure}
    \centering
    \includegraphics[width=\linewidth]{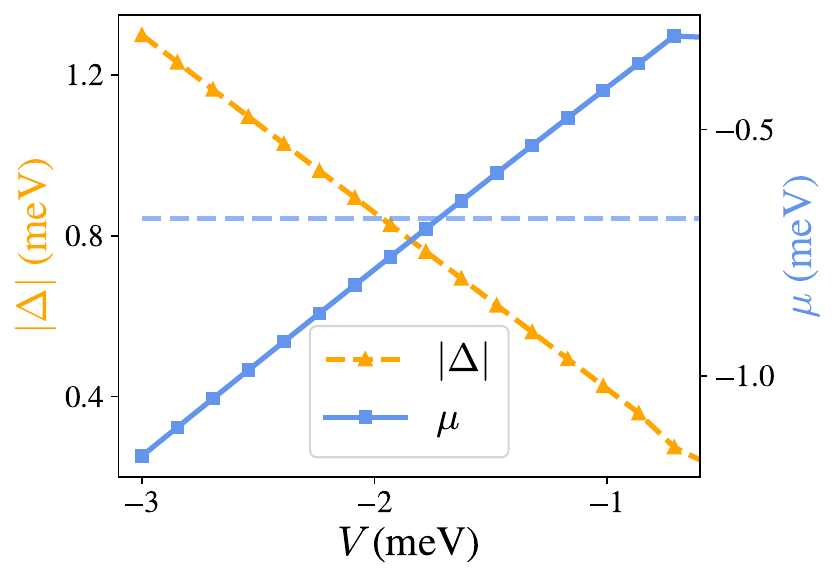}
\caption{Order-parameter amplitude $|\Delta|$ (left axis) and chemical potential $\mu$ (right axis) versus coupling strength $V$. The dashed line marks the flat-band bottom. For attractive $V \approx -1.7\,\mathrm{meV}$ ($\Delta\simeq 0.73  $meV), the system crosses into the BEC-like regime.    }

    \label{fig5}
\end{figure}

It is useful to make a few remarks on intervalley pairing. From Figs.~\ref{fig3} and \ref{fig5}, one can directly read off that the attractive interaction required in our scenario corresponds to an energy scale on the order of $1$meV. This scale should be readily accessible experimentally, and it is notably smaller than the $\sim 0.1$~eV estimate reported in Ref.~\cite{Julku2020} for intervalley pairing in a nearest-neighbor attraction model. Within the class of short-range (non-local) models considered here, we thus expect intravalley pairing to be a more favorable ground-state candidate than intervalley pairing.

To summarize, the stable superconducting order, which corresponds to the global minimum of the potential $\Omega(\mu, {\bf Q}, \Delta_L, {\bf S})$ for the parameters considered, is characterized by equal condensates in the two layers, a pairing momentum at the $M$ point, and unitary spin-triplet pairing.

\subsection{Nematic Nature of Superconducting order}\label{Nematic} 
The superconducting order considered here is \emph{nematic}. Note that this anisotropy does \emph{not} originate through a 
pre-determined pairing interaction,
but rather a consequence of condensation. In particular, the ground state condenses at a $M$ point in the Brillouin zone.  Since there are three symmetry-related $M$ points, selecting one spontaneously breaks the threefold rotational symmetry $C_3$. For completeness, we also provide a proof for an arbitrary pairing momentum $\mathbf{Q}$, see the Methods section, “Proof of $C_3$-SSB.”

  This is in contrast to the inter-valley case, where nematicity will emerge after selecting a particular combination of the $E$-channels\cite{PhysRevResearch.2.033062,Christos2023}.

\begin{figure*}
\includegraphics[width=6.5in]{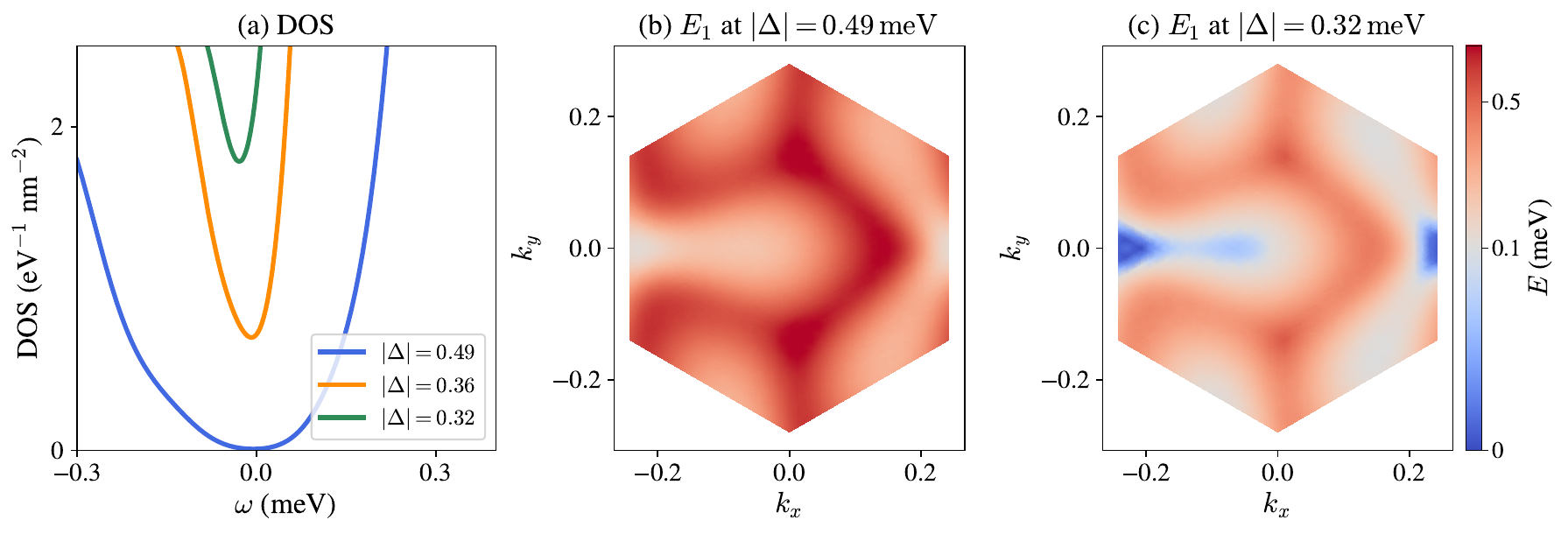}
\caption{(a) Density of states (DOS) versus energy $\omega$ (meV) for three 
different gap amplitudes, with curves labeled by $|\Delta| \simeq 0.49, 
0.36,$ and $0.32$ meV. As the order parameter is reduced, the bottom of the characteristic U-shaped gap narrows towards a V-shape.  Notably, a finite zero-bias conductance emerges in the latter case 
(b) This shows the lowest positive Bogoliubov-de Gennes eigenvalue $E_1(M, \mathbf{k})$ 
plotted across the mini-Brillouin zone for strong attraction, $|\Delta| = 0.49$ meV. At this amplitude, the spectrum remains 
fully gapped, corresponding to the U-shaped DOS. 
{(c).}  At weaker attraction and thus a smaller amplitude for $|\Delta| = 0.32$ meV, a small Bogoliubov 
Fermi surface appears, as should be evident in the figure, 
and the (V-shaped) DOS spectrum becomes gapless.
}
\label{fig6}
\end{figure*}

\section{ U to V-like transitions in tunneling}

In the trilayer case, where Kekul\'e superconductivity is similarly observed~\cite{Kim2023}
the evolution of the conductance with doping is of particular interest.
The tunneling spectrum which reflects the
quasi-particle density of states~\cite{Kim2022}, 
has a V-like shape (often with a ``zero bias conductance")
when filling is in the regime closer to $\nu = -3$,
while it transitions to a U-shaped spectrum, when $\nu$ is nearer
$-0.2$. 
It is generally believed that these spectra, then, reflect
a transition from a gapped superconductor to a nodal superconductor.
Important to this analysis, one can associate
the U regime with a shorter Landau Ginsberg coherence length which 
reflects~\cite{Chen2024} stronger attractive interactions.
In this context, while originally this U transition was thought to relate to a  
BEC regime, there are claims now to refute this speculation \cite{Van2025}.

Although these experiments address the tri-layer case, it is useful
here to study the bilayer density of states (DOS), using the parameter set corresponding to Fig.~\ref{fig2}. As shown in Fig.~\ref{fig6}(a), for a superconducting order parameter of $|\Delta| \geq 0.4$ meV, the DOS exhibits a clear U-shape, indicating that the quasiparticle excitations are fully gapped. 

As the order parameter is reduced, the flat bottom of this U-shaped gap narrows, eventually transitioning into a smoothed V-shape with a finite DOS at zero frequency. 
We find this corresponds
to the fact that there are touching points leading to a very small Bogoliubov
Fermi surface contained within the quasi-particle dispersions for the
``V case" and missing for the ``U case", as can be seen in
Fig.~\ref{fig6}(b) and
Fig.~\ref{fig6}(c) \footnote{Although their Bogoliubov Fermi surfaces are larger,
there is some similarity here to work from
Christos et al, Nat. Commun. 14, 7134 and Agterberg et al, Phys. Rev. Lett. 118, 127001}. We emphasize here
that there is generally a finite conductance at zero energy (zero bias
conductance) when the
V shape is present.  This constitutes a prediction from the present theory.

To address this in more detail, it is useful to
consider the Bogoliubov dispersion in the two-flat-band limit:
\begin{equation}
E(M,{\bf q}) = \delta \epsilon({M,\bf q}) \pm \sqrt{\bar{\epsilon}({M,\bf q})^2 + \Delta^2 \varphi({M,\bf q})^2}\label{BdG_off}
\end{equation}
where $\delta \epsilon({M,\bf q}) = [\epsilon_1(M+{\bf q}) - \epsilon_2(M-{\bf q})]/2$ and $\bar{\epsilon}({M,\bf q}) = [\epsilon_1(M+{\bf q}) + \epsilon_2(M-{\bf q})]/2$ are derived from the dispersion of two flat bands, $\epsilon_{1,2}({M,\bf q})$. Here $\varphi(M,q)=\sum_{L}\Lambda^t_{12,L}(M,q)$ represents the strength of off-diagonal pairing\footnote{From Fig.~2, we observe that the off-diagonal components of $\Lambda$ are significantly larger than the diagonal ones. Consequently, we retain only the off-diagonal pairing.}.

From this expression, two regimes emerge. When $\Delta$ is large enough such that the condition $|\Delta \varphi(M,{\bf q})| > |\delta \epsilon(M,{\bf q})|$ holds for all ${\bf q}$, the dispersion is necessarily gapped, leading to the U-shaped DOS. Conversely, when $\Delta$ is small enough that $|\Delta \varphi(M,{\bf q})| \leq |\delta \epsilon(M,{\bf q})|$ for some ${\bf q}$, gapless points can develop. This results in a V-shaped DOS generally
associated with a small Bogoliubov Fermi surface, reflecting the sharpness of the V.
We anticipate the existence of a critical value, $\Delta_c$, that marks a transition from a fully gapped to a gapless superconducting state.

We emphasize that the qualitative BFS signature---a $V$-shaped tunneling DOS with finite zero-bias conductance---is expected to be robust under realistic conditions. Finite temperature and quasiparticle lifetime primarily broaden the spectrum (which can be modeled by a phenomenological broadening of the DOS), modifying the low-energy DOS quantitatively but not eliminating the finite  weight. Weak, momentum-conserving intervalley mixing generically deforms the zero-energy contour rather than instantly gapping it, except at special overlap points. Moderate disorder similarly acts through lifetime broadening and pair breaking. In our mechanism, reducing $|\Delta|$ tends to expand the regime where $|\delta\epsilon|$ exceeds the local gap. A more detailed discussion and modeling of these effects is provided in Methods section ``Robustness of the Bogoliubov Fermi surface".

It is also useful to contrast our BFS-based $V$-shaped spectrum with other recent proposals. In Ref.~\cite{PhysRevLett.133.146001} and Ref.~\cite{PhysRevB.110.045133} , the $V$-shape is associated with a nodal superconducting state, with a low-energy spectrum characterized by point nodes. In these nodal scenarios, the low-energy DOS vanishes, so the zero-bias conductance is expected to approach zero. By contrast, in our theory the $V$-shaped spectrum in the small-$|\Delta|$ regime originates from a Bogoliubov Fermi surface, which yields an \emph{intrinsic} finite zero-energy DOS already in the clean limit and therefore a \emph{finite} zero-bias conductance. This qualitative distinction provides a direct experimental discriminator between the nodal and the BFS mechanisms. See Fig.~\ref{fig7}

\section{Zero-bias Conductance}

\begin{figure}
\includegraphics[width=3.5in]{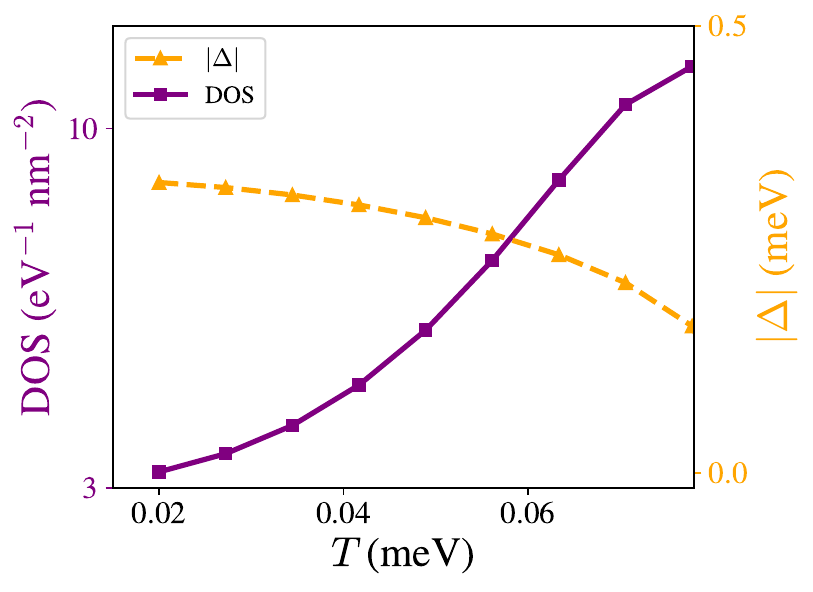}
\caption{Density of states and order parameter versus temperature, computed at the
interaction strength
$V=
-0.8
$meV. As temperature increases, the order parameter decreases while the DOS
increases. This behavior is consistent with the presence of a Bogoliubov Fermi surface
and can be seen to be related to Figure 3e in Ref.~\cite{Park2025} (see text), which addresses the zero-bias conductance. }
\label{fig7}
\end{figure}

We quantify the calculated zero-bias conductance in tunneling spectra as a function of 
temperature in Fig.~\ref{fig7}. 
The fact that
 two different groups
\cite{Park2025,Kim2022}
have
arrived at quite similar plots suggests that this may be an intrinsic
phenomenon.
What is important here is that the zero bias conductance, in principle,
contains information about the excitation
spectrum of the superconductor. Moreover, there should be some distinction
between the U and V shaped cases which will be important to assess in future,
although the focus of the experiments thus far has been in the V-shaped regime.

We present a plot of the zero bias conductance in this regime,
as a function of temperature
which is compared with the temperature dependence of order parameter $\Delta$.
The plot in
Fig.~\ref{fig7} is inspired by a related plot (see their Fig 3e) for the 
moir{\' e} materials in Ref.~\cite{Park2025}.
In the Methods section we discuss the U-shaped counterpart.

One can speculate that here the V-shaped tunneling spectra may be qualitatively different from the classic
$d$-wave behavior
in the cuprates~\cite{Pushp2009} where a plot 
$dI/dV \propto V$  vanishes at zero bias.
Throughout the cuprate family, however, there are some indications of
a finite zero bias conductance but these are generally understood~\cite{Reber2013} as associated
with a Dynes lifetime parameter in the
tunneling density of states.
We argue that what is different about the
moir{\'e} systems is that the finite
zero-bias conductance appears order-parameter controlled, not merely
lifetime-broadened. This suggests a superconducting state whose excitation
spectrum remains partially gapless due to intrinsic features of the paired state itself.
Given the consistency across a few fillings~\footnote{Private Communications with Hyunjin Kim} and across groups~\cite{Park2025,Kim2022}, we thus attribute the finite zero-bias conductance to an intrinsic Bogoliubov Fermi surface rather than to extrinsic, random Dynes-like broadening.

\section{Overview of Pairing Theories }
We have discussed the robustness of our results under reasonable variations of the model parameters, including the kinetic bandwidth within the BM model and the interaction range. Here we further explain why it is appropriate to use the BM dispersion and provide additional discussion of the attractive interaction.

We have emphasized the important role of screening in TBG, which renders the repulsive core of the Coulomb interaction effectively short-ranged. Superconductivity that arises indirectly from repulsive interactions is most systematically treated within Eliashberg theory~\cite{Eliashberg1963,Levin1979a,Scalapino2012}. A crucial feature of this theory is that it is guaranteed to satisfy conservation laws.
As a consequence this framework requires the use of the same screened interaction, but with different projections, to address the bandstructure renormalizations and the pairing channel. 

For example in $p$-wave paired superfluid-$^3$He or in the $d$-wave paired cuprates the projections in the bandstructure channel are
effectively isotropic while the attractive pairing channel reflects the appropriate $p$ or $d$-wave anisotropy.
Comparison with the superfluid-$^3$He literature~\cite{Levin1979a,ZQ2025} indicates that, provided there are no degeneracies among candidate pairing states~\cite{Anderson1973}, the pairing symmetry can be identified reliably even within a simplified treatment that neglects detailed band-structure renormalization
~\cite{Anderson1973}. By contrast, self-energy corrections are important for obtaining quantitative estimates 
of $T_c$~\cite{Levin1979a,ZQ2025}. Since our focus here is the qualitative structure of the superconducting order at low temperature, we adopt this simplified procedure~\cite{Anderson1973} for moir\'e graphene.

There is, of course, an ongoing debate about whether a more conventional electron--phonon mechanism is appropriate for the twisted graphene family. This viewpoint is supported by an extensive literature~\cite{Chen2024,Kwan2024,Shi2025,Wang2024,Liu2024}. We argue here that the strong pairing, as evidenced by the short coherence lengths~\cite{Cao2018,Park2021} in these materials, is suggestive of an electronic mechanism. Moreover, the pairing is generally reported~\cite{Park2026,Banerjee2025} to be non-$s$-wave, and the bands are sufficiently flat that retardation is compromised; both observations support pairing arising from a repulsive (e.g., Coulomb) interaction. We emphasize that it remains premature to rule out any pairing mechanism at present.

In recent experiments~\cite{Stepanov2020,Saito2020}, the authors have sought to disentangle correlated-insulator behavior at integer fillings from superconductivity, and claim that this argues against fluctuation-mediated pairing tied to correlated insulating states in favor of a more conventional electron--phonon mechanism. We point out here~\cite{ZQ2025} that reducing the strength of the Coulomb interaction can suppress competing orders that tend to preempt superconductivity. 
More specifically, in the large class of magnetically proximate superconductors,
the strongest manifestations of this superconducting pairing are generally correlated with the
weakest magnetism.
As a consequence, when magnetic order is absent or diminished in a particular region of phase space, the remnant fluctuations
enable the expansion of magnetism-driven superconductivity into that very region
~\cite{Landaeta2018}. This prior analysis of non-phononic superconductivity is in line with what is observed
~\cite{Stepanov2020,Saito2020} in the twisted graphene family.

\section{Discussion and Relation to Prior Work}
This paper studies a Kekul\'e superconducting order in the twisted graphene family, corresponding to an intravalley pairing state living on the AB bond. The consequences of this simple ansatz are consistent with a range of existing experimental phenomenology, including the Kekul\'e pattern in STM~\cite{Nuckolls2023,Kim2023}, non-singlet pairing inferred from Pauli-limit violation~\cite{Cao2021}, nematicity suggested by transport~\cite{caoNematic}, the $U$-to-$V$ evolution of the tunneling DOS~\cite{Kim2022}, and the systematic behavior of the zero-bias conductance~\cite{Kim2022,Park2026}. Underlying these latter two observations is the emergence of a Bogoliubov Fermi surface (BFS), which also implies a $T^2$ scaling of the superfluid stiffness~\cite{Tanaka2025,Wang2026Superfluid}, as expected from a Sommerfeld expansion~\cite{Wang2026}. Importantly, these observations arise in concert with a prediction: for relatively strain-free samples, the PDW order induces a nonuniform charge modulation at moir\'e wavevectors near the $M$ point, which should be observable in STM.

Furthermore, we have verified that the qualitative superconducting features discussed above remain robust over a reasonable range of twist angles (corresponding to flat-band bandwidths from $1$ to $10$~meV) and interaction ranges (up to $\sim 10$ graphene lattice constants). By contrast, quantitative results, such as the dependence of the order parameter on interaction strength and the relationship between the critical temperature $T_c$ and $\Delta$~\cite{Wang2026Superfluid}, are model-dependent and will vary with microscopic details.


There have been many different theoretical ideas about the superconductivity proposed in the literature for graphene-based
materials
\cite{You2019,Wu2018,Lian2019,Cea2021,Hu2019,Julku2020,Christos2023,Khalaf2021,Gonzalez2019,Po2018,PhysRevLett.133.146001,PhysRevB.111.174523,PhysRevX.10.031034,aat4698}.
With the more recent
observations of Kekul\'e order for the twisted graphene superconductors \cite{Nuckolls2023,Kim2023}
(which has not been addressed in this
body of theoretical work), one needs now to incorporate
such effects. They will enter either through the pairing
glue or in the
formal machinery.

At the heart of our paper
is an observed correlation~\cite{Kim2023} in these moir{\' e systems between
the highest Kekul\'e peaks found to be associated with the strongest pairing regime.
Thus, we argue against the interpretation that Kekul\'e order acts as a pairing "glue". We know
in superconductors which have a non-phononic, bosonic (e.g., magnetic)  pairing mechanism, the superconductivity is
strongest when the non-superconducting, alternative ordering is weakest\cite{ZQ2025}.
Rather Kekul\'e order in this twisted graphene system seems to be intertwined with the pairing machinery as distinct from
the pairing
mechanism.
This order then leads to a PDW-like description of the superconductivity\cite{Tsuchiya2016,Roy2010,PhysRevLett.114.237001,Wang2026}.

While the importance of a Bogoliubov Fermi surface (BFS) for understanding the $U$-to-$V$ transition was emphasized in the pioneering work of Christos, Sachdev, and Scheurer~\cite{Christos2023}, we stress that the origin and symmetry protection of the BFS in their work differ from that
discussed here. In Ref.~\cite{Christos2023}, $C_{2z}$ symmetry and a spin-polarized triplet state inherited from the normal phase play a central role in protecting the BFS. By contrast, we make no presumptions about the spin symmetry in the pairing,
and $C_{2z}$ symmetry is absent in a single-valley pairing process. Rather, finite-momentum pairing produces a ``mismatch'' that prevents a fully gapped spectrum and yields gapless regions. The resulting BFS is reasonably robust within its phase so that
the $U$-to-$V$ crossover with decreasing attraction and the associated increased zero-bias conductance should be viewed as intrinsic~\footnote{We have checked different pairing momenta $\mathbf{Q}$ across the mini Brillouin zone and both spin-singlet and spin-triplet channels; these two qualitative features persist throughout.}.

We have, throughout this paper, made few distinctions between the bi- and tri-layer
moir\'e graphene superconductors.
We believe, in line with most of the experimental phenomenology,
that many of the general features discussed here apply to both as we focus
almost exclusively on the physics of the two flat bands (with a few nearby remote bands) and the related symmetries. There is an interesting distinction
however which is that the attractive interactions (as measured by the large size of the pairing
gap and the very short Landau Ginsberg coherence lengths) may not be strong enough in the bilayer case to reveal a U shaped
tunneling spectra. The V shape tunneling along with the systematic zero bias conductance should remain
and this seems to be currently consistent with experiments on the bilayer superconductors\cite{Oh2021}.

\appendix

\section{Methods}

\subsection{Bistritzer--MacDonald Model and Symmetries} 

We briefly review the Bistritzer--MacDonald (BM) model of twisted bilayer graphene (TBG) and its symmetries. When two layers are stacked with a relative twist angle $\theta$, a fundamental momentum scale emerges,
\begin{equation}
    k_\theta = 2K \sin \frac{\theta}{2}, 
    \qquad K = \frac{4\pi}{3\sqrt{3}a},
\end{equation}
where $a$ is the graphene lattice constant. A convenient choice of momentum difference between the two Dirac points is
\begin{equation}
    \mathbf{q}_1 = \mathbf{K}_{\text{top}} - \mathbf{K}_{\text{bot}} = (0,1)\,k_\theta .
\end{equation}
The remaining moiré reciprocal vectors can be defined as $\mathbf{b}_{1,2} = \mathbf{q}_{2,3} - \mathbf{q}_1$, where $\mathbf{q}_{2,3}$ are obtained by counterclockwise rotations of $\mathbf{q}_1$ by $120^\circ$.  

The BM Hamiltonian describes the continuum theory near the two Dirac cones,
\begin{equation}
    H = v\Big(-i\nabla - L_z\mathbf{q}_1/2 + \mathbf{M}\Big)\cdot \boldsymbol{\sigma}_{\theta/2} 
    + \big[T(\mathbf{r})L^+ + \text{h.c.}\big],
\end{equation}
where $v$ is the Fermi velocity of monolayer graphene, $\mathbf{M}$ denotes the $M$-point of the mini Brillouin zone, $L$ and $\sigma$ act on layer and sublattice indices, and $\boldsymbol{\sigma}_{\pm\theta/2}$ denote Pauli matrices rotated by angle $\pm \theta/2$. The tunneling matrix $T(\mathbf{r})$ is parameterized by two amplitudes $w_{AA}$ and $w_{AB}$,
\begin{equation}
    T(\mathbf{r}) = w_{AA}\,\alpha_0(\mathbf{r})
    + w_{AB}\,\sigma^+ \alpha_1(\mathbf{r})
    + w_{AB}\,\sigma^- \alpha_2(\mathbf{r}),
\end{equation}
with spatial form factors
\begin{equation}
    \alpha_m(\mathbf{r}) = \sum_{j=0}^2 w^{mj}\,e^{-i\mathbf{b}_j\cdot\mathbf{r}}, 
    \qquad b_0={\bf 0},\quad w = e^{i2\pi/3}.
\end{equation}

The BM Hamiltonian respects four fundamental symmetries: moiré translations, threefold rotations $C_3$, mirror reflection $M_y$, and the combined twofold rotation with time reversal $C_2\mathcal{T}$. For a spatial transformation $R$, the Hamiltonian transforms as
\begin{equation}
    H'(\mathbf{x}) = H(R^{-1}\mathbf{x}),
\end{equation}
and invariance under the symmetry requires
\begin{equation}
    H'(\mathbf{x}) = U(\mathbf{x}) H(\mathbf{x}) U^\dagger(\mathbf{x}),
\end{equation}
with $U(\mathbf{x})$ a unitary operator. Explicitly, the mirror symmetry and threefold rotation\cite{PhysRevB.99.035111} are represented by $U_{M_y} = \sigma_x L_x$ and
\begin{equation}
    U_{C_3}(\mathbf{x}) = \exp\!\left(\tfrac{2\pi i}{3}\,\hat{z}\cdot\boldsymbol{\sigma}\right) 
    \exp\!\left[i\Big(2\mathbf{M} + (L_z+1)\mathbf{b}_1/2\Big)\cdot \mathbf{x}\right].\label{22}
\end{equation}
For the antiunitary $C_2\mathcal{T}$ operation, one has
\begin{equation}
    H'(\mathbf{x}) = H^*(-\mathbf{x}), \qquad 
    H'(\mathbf{x}) = \sigma_x H(\mathbf{x}) \sigma_x,
\end{equation}
which guarantees invariance under the combined symmetry.

\subsection{Flat band Approximation}
We find that \(\Lambda\) is highly nonlocal in the band index: sizeable matrix elements \(|\Lambda_{mn}(\mathbf{q})|\) persist even for \(|m-n|\gg 1\). Thus it is necessary to investigate the coupling between the flat bands and remote bands. Experimental data and numerical simulations consistently indicate a significant separation of energy scales. The superconducting gap is typically on the order of $\Delta\sim 1~\mathrm{meV}$, whereas the single-particle gap separating the flat and remote bands in the 
BM model is much larger, at $E_{\text{band}}\sim 50~\mathrm{meV}$. A Schrieffer–Wolff transformation can be employed to integrate out the remote bands:
\begin{equation}
\tilde{H}_{LL}\simeq H_{LL}+H_{LH}\,(-H_{HH})^{-1}H_{HL}.
\end{equation}
Here, $H_{LL}$ is the Hamiltonian in the flat-band subspace, $H_{LH}$ couples the flat and remote bands, and $H_{HH}$ is the Hamiltonian within the remote-band subspace.
 After the transformation, the new Hamiltonian $\tilde{H}_{LL}$ acquires a correction of order $\Delta^2/E_{\text{band}}$.
When this energy is much smaller than the bandwidth \(W\) of a {\it single} flat band, the isolated-flat-band description of intra-valley superconductivity is well-justified, provided the bands are not
perfectly flat (\(W\!\to\!0\)) or \(\Delta\) sufficiently large so that remote-band effects must be included.

In numerical simulations that incorporate remote bands into the gap equation, we find that including only the two flat bands yields reasonable results, although adding more bands enhances precision. In practice, we retain $N=20$ bands.

\begin{figure}
    \centering
    \includegraphics[width=\linewidth]{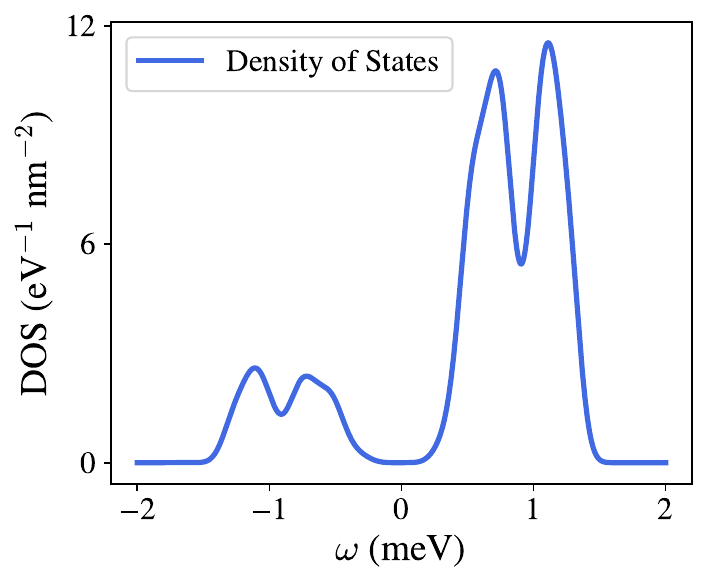}
    \caption{ U-shaped density of states (DOS) from two-flat-band superconductivity.
The figure is plotted at $|\Delta|\simeq 0.63~\mathrm{meV}$ ($V\simeq -1.5$ meV). The  dip feature around $\omega\sim$ 1meV originates from the band touching of two BM flat bands at $K/K'$ points.
    }
    \label{appendixfig1}
\end{figure}

\begin{figure}
    \centering
    \includegraphics[width=\linewidth]{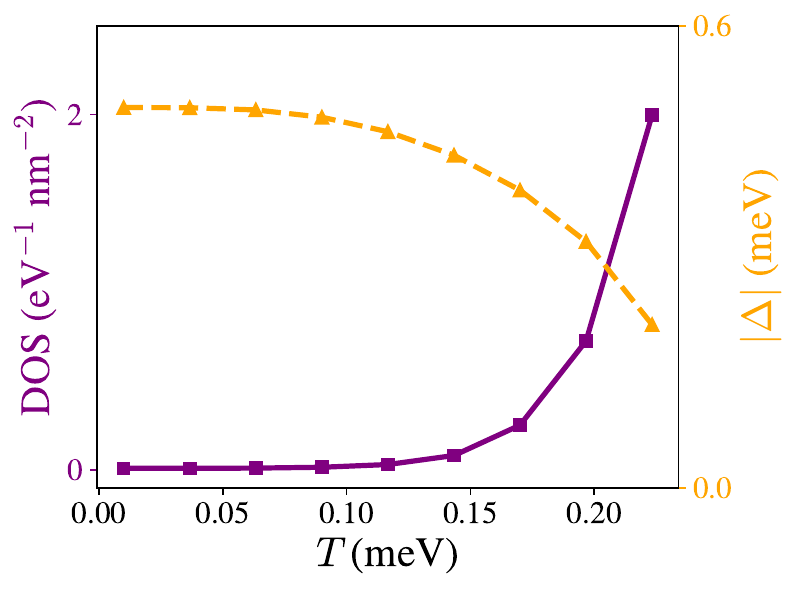}
    \caption{
Quasi-particle density of states (reflecting zero bias conductance) 
and order parameter versus temperature, computed at the
interaction strength
$V\simeq
-1.17
$meV. This is in the U-shaped tunneling regime and should be contrasted with
Fig.~\ref{fig7} 
discussed earlier which is for the V-shaped case with weaker attractive interaction.
As temperature increases, the density of states remains gapped until the order parameter
$\Delta$ drops sufficiently and then the system crosses over to a more V-like spectrum associated
with a
Bogoliubov Fermi surface.} 
\label{appendixfig2}
\end{figure}

\section{U/V-transitions in the DOS}
As discussed in the main text, the U/V transition can be understood
through the analysis presented in Eq.~(\ref{BdG_off}). For a typical
order parameter, say $\Delta \sim 0.3$ meV, the associated Bogoliubov Fermi
surface is small, as depicted in Fig.~\ref{fig6}. This implies that the ``V" shape has a rather sharp minimum.

Here, we present the full plot of the density of states (DOS) within the
U-shaped regime in Fig.~\ref{appendixfig1}. While the low-energy behavior
has already been presented, substructure corresponding to dips are observable around $\omega \sim 1$ meV.
These dips arise from the structure of two flat bands that touch at the $K$
and $K'$ points.  The particle-hole asymmetry in
the DOS is a result of our choice of filling which is taken to be $3/8$ of the
lower flat band (equivalent to $\nu=-2.5$ filling in experiments).

In Fig.~\ref{appendixfig2}
we plot the effective zero bias conductance as well as the order parameter for a strong
attraction case where the tunneling curve assumes a U-shape. This represents a 
prediction as, thus far, there are no published curves for this zero bias
conductance.
In contrast to 
Fig.~\ref{fig7} at weaker attraction, here for stronger attraction, as temperature 
increases the zero bias density of states remains gapped until crossing over to the behavior
associated with the V-shaped regime where a Bogoliubov Fermi surface is present.

 \subsection{Proof of $C_3$-SSB}
Here we provide a proof that the intravalley superconducting order breaks $C_3$-symmetry. In the BM model, there is a $C_3$ symmetry about $\Gamma$. A counterclockwise $C_3$ rotation is represented by the unitary operator in Eq.~\ref{22}.

The action of \(U_{C_3}\) on the pairing operator \(\hat{\Delta}(\mathbf{r})\) induces a layer-dependent momentum boost:
\begin{equation}
    \label{17}\hat{\Delta}_{L,\sigma\sigma'}(\mathbf{r}) \;\longmapsto\;
    e^{\,2 i\,\mathbf{p}_L\!\cdot \mathbf{r}}\;\hat{\Delta}_{L,\sigma\sigma'}(\mathbf{r}) .
\end{equation}
Here \(\hat{\Delta}(\mathbf{r})\) denotes the operator corresponding to Eq.~\eqref{3}, and only the layer index $L$ is preserved for simplicity, and
\(\mathbf{p}_L = 2\mathbf{M}\) for the bottom layer while
\(\mathbf{p}_L = 2\mathbf{M} +  \mathbf{b}_1\) for the top layer.
This phase cannot be removed by a global \(\mathrm{U}(1)\) gauge choice: it is a \emph{layer-dependent}
linear function of \(\mathbf{r}\) (a boost), so gauging it away in one layer necessarily induces an incompatible shift in the other.

First consider condensation at the highest-symmetry $ \Gamma$ point of the model and take this as the reference momentum, i.e., the pairing momentum \(\mathbf{Q}=0\). Here we examine the symmetry of mean-field Hamiltonian, i.e., fixing condensate and transforming fermions\footnote{An equivalent approach is to show that the condensate is not invariant under 
$C_3$. The proof is similar.}.  Since the BM-model term is invariant under a \(C_3\) rotation, we only need to examine the mean-field term:
\begin{equation}
-\sum_{\mathbf{q},m,n}\Delta_{L,\sigma\sigma'}\,
\Lambda^*_{mn,L}(\mathbf{0},\mathbf{q})\,
\hat{c}^\dagger_{\sigma,m}(\mathbf{q})\,
\hat{c}^\dagger_{\sigma',n}(-\mathbf{q}) .
\end{equation}
A \(C_3\) rotation acts as 
\begin{equation}
    \Lambda^*_{mn,L}(\mathbf{0},\mathbf{q})
    \;\longmapsto\;
    \Lambda^*_{mn,L}(-\mathbf{p}_L,\mathbf{q}),
    \label{LambdaShift}
\end{equation}
which shifts the pairs’ center-of-mass momentum by \(\mathbf{p}_L\).
Recalling the definition in Eq.~\ref{5}, the Bloch-amplitude product \(u_{\mathbf{G}}u_{-\mathbf{G}}\) in the form factor becomes
\(u_{\mathbf{G}-\mathbf{p}_L}u_{-\mathbf{G}-\mathbf{p}_L}\), thereby modifying the form factor.
Hence the Bogoliubov spectrum is not invariant under \(C_3\), which is consistent with the form factor presented in Fig.~\ref{fig2}. 

If the condensate forms at a finite momentum \(\mathbf{k}\neq\Gamma\), \(C_3\) is likewise broken: \(\mathbf{k}\) itself
rotates under \(C_3\), and the shift in Eq.~\eqref{LambdaShift} still applies. This completes the proof that superconducting order
generically breaks threefold rotational symmetry. Importantly, the argument does not rely on any particular form of the attractive
interaction; it uses only that the superconducting order is intra-valley.

  \subsection{Robustness of the Bogoliubov Fermi surface}
\label{sec:BFSrobust}

In this section, we provide further details regarding the stability of the Bogoliubov Fermi Surface (BFS) and the resulting density of states (DOS) signatures against realistic experimental perturbations, specifically addressing finite temperature, intervalley mixing, and disorder.

Regarding finite temperature, two effects are particularly relevant. First, thermal and lifetime broadening in the tunneling DOS can be modeled phenomenologically by replacing the $\delta$-function with a broadened kernel of width $\eta$. We have verified that the BFS contribution to the DOS is robust under such broadening: the zero-energy DOS remains finite for $\eta > 0$, and its magnitude changes only perturbatively for small $\eta$. In particular, the qualitative $V$-shaped spectrum with finite zero-bias conductance persists. Second, the temperature dependence of the self-consistent solution modifies the spectrum quantitatively as temperature increases. The order parameter $\Delta(T)$ is obtained from the finite-$T$ gap equation, and we present this evolution in the Fig.~\ref{fig7}. Importantly, the experimentally observed zero-bias conductance exhibits a similar temperature dependence.

The stability of the BFS is also maintained in the presence of weak intervalley mixing. In the moir\'e-periodic (crystal-momentum-conserving) case, intervalley hybridization couples states at the same moir\'e momentum, $(K, \mathbf{k}) \leftrightarrow (K', \mathbf{k})$, whereas time reversal relates $(K, \mathbf{k})$ to $(K', -\mathbf{k})$. As a result, a zero-energy state on the $K$-valley BFS at momentum $\mathbf{k}$ is not generically degenerate with a zero-energy state in the opposite valley at the same $\mathbf{k}$. In this generic situation, turning on a small intervalley hybridization shifts the low-energy eigenvalues smoothly with the mixing strength, allowing the BFS contour to deform continuously rather than being immediately gapped. 

We do raise the caution that while a gap can open if the two valley zero-energy contours overlap at the same $\mathbf{k}$, there is no symmetry that enforces a $\mathbf{k} \to -\mathbf{k}$ constraint within a single valley. Consequently, an exact overlap is not expected except at special points. Away from these points, weak momentum-conserving intervalley mixing primarily renormalizes the dispersion and leaves the BFS intact.

Finally, we distinguish between two common effects of disorder. Quasiparticle lifetime broadening smears the DOS in a manner similar to finite-$T$ broadening; we have verified that the qualitative $U$-to-$V$ crossover and the finite low-energy DOS persist under reasonable broadening. Additionally, pair-breaking disorder can reduce the magnitude of $|\Delta|$. Because the BFS appears when the effective asymmetry exceeds the local gap, reducing $|\Delta|$ tends to enlarge the parameter regime over which a BFS occurs, provided the disorder is not so strong as to suppress superconductivity altogether. Therefore, we find that the qualitative DOS signatures remain robust in the presence of finite temperature, weak intervalley mixing, and moderate disorder.

\section{Code's functionality}

All numerical results reported in this work were produced using a reproducible Python/Jupyter workflow. The starting point is an implementation of the Bistritzer--MacDonald (BM) continuum model for twisted bilayer graphene with tunable parameters $(w_0,w_1)$ and twist angle $\theta$. From the BM Bloch eigenstates evaluated on a momentum grid in the moir\'e Brillouin zone, we construct the particle--particle form factors (the $\Lambda$-matrix) that enter the pairing problem. This step is carried out in the notebook \texttt{Lambda\_Matrix\_Generate.ipynb}. In practice, the BM parameters are set in the block labeled \texttt{TBG\_Parameter}, after which the notebook computes the band energies and Bloch wavefunctions; it then fixes a pair center-of-mass momentum $Q$ (for example, at the moir\'e $M$ point) and evaluates/stores the corresponding form factors in the block labeled \texttt{COM\_Tune}.

To resolve pairing in different symmetry sectors, we project the raw form factors into symmetry channels separately for singlet and triplet pairing. This symmetry-channel construction is performed using \texttt{read\_Lambda\_Construct\_singlet.ipynb} (singlet) and \texttt{read\_Lambda\_Construct\_triplet.ipynb} (triplet). By default, these notebooks generate the fully symmetric $A$ channel (\texttt{sym=0}); additional channels (e.g., $E_1$ and $E_2$) are obtained by updating the symmetry indices in the block labeled \texttt{loc\_sym} (for example, setting \texttt{syms=(0,1,2)}). The resulting symmetry-resolved $\Lambda$-matrices are saved and used as inputs to the subsequent self-consistent calculations.

Given the stored singlet/triplet form factors in a chosen symmetry channel, we solve the superconducting gap equation either in the grand-canonical or canonical ensemble. The grand-canonical solver is implemented in \texttt{Solvegrand\_singlet.ipynb}; the pairing sector is selected by setting \texttt{flat\_dir} to the appropriate dataset folder (e.g., \texttt{"lambda\_singlet"} or \texttt{"lambda\_triplet"}), and the main control parameters include the temperature $T$, the interaction strength, and the subspace dimension used for numerical projection. The canonical-ensemble solver is implemented in \texttt{Solve\_singlet.ipynb}. In both cases, the converged solutions and associated spectral quantities are written to disk and subsequently analyzed for the figures.

The main-text figures are reproduced by notebooks in \texttt{/figures}, which read the stored outputs from the above steps. Figure~2 is generated by \texttt{fig2.ipynb}, which plots the BM band dispersion and representative form-factor information using outputs from \texttt{Lambda\_Matrix\_Generate.ipynb} together with the triplet-channel construction. Figure~3 is generated by \texttt{fig3.ipynb} and compares singlet and triplet solutions using baseline datasets and/or datasets produced by the grand-canonical solvers. Figure~4 is generated by \texttt{fig4.ipynb} using canonical-ensemble results. Figure~5 is generated by \texttt{fig5.ipynb}, which produces tunneling density-of-states plots and representative Bogoliubov--de Gennes dispersions from the converged solutions. Figure~6 is generated in two steps: \texttt{preparation\_fig6.ipynb} preprocesses the solver outputs to compute the zero-bias conductance datasets, and \texttt{fig6.ipynb} produces the final plots.

\section{Data availability}
The data analyzed in the current study are available from the author Ke Wang on reasonable request.

\section{Code availability}
The code used in this study is available at https://github.com/KernelW/arXiv-2510.06451.

\vskip5mm

\section{Acknowledgement}
We thank Kevin Nuckolls, Hyunjin Kim, Ivar Martin, Bitan Roy, Stevan Nadj-Perge, Senthil Todadri, Zhiqiang Wang,
Ananth Malladi for helpful discussions. We also acknowledge
the University of Chicago’s Research Computing Center for
their support of this work.

\section{Author Contributions}
K.L. conceived and supervised the project. K.W. performed the computations. K.W. contributed to the acquisition of the data and preparation of figures. All authors have contributed to the interpretation of the data and the drafting as well as the revision of the manuscript.

\section{Competing Interests}
The authors declare no competing interests.

\section{ ADDITIONAL INFORMATION}
Correspondence and requests for materials should be addressed to the authors K. Wang and K. Levin.

\bibliography{References}

\end{document}